\documentclass[conference]{IEEEtran}
\IEEEoverridecommandlockouts
\usepackage{cite}
\usepackage{amsmath,amssymb,amsfonts}
\usepackage{algorithmic}
\usepackage{graphicx}
\usepackage{textcomp}
\usepackage{xcolor}
\usepackage{hyperref}
\usepackage{booktabs}
\usepackage{multirow}

\def\BibTeX{{\rm B\kern-.05em{\sc i\kern-.025em b}\kern-.08em
    T\kern-.1667em\lower.7ex\hbox{E}\kern-.125emX}}
\begin{document}

\title{IUMENTA*: A generic framework for animal digital twins within the Open Digital Twin Platform
\thanks{\footnotesize \textsuperscript{*}Note: Latin for livestock.}
}

\author{
\IEEEauthorblockN{Ali Youssef$^1$, Kristina Vodorezova$^1$, Yannick Aarts$^1$, Wisdom E. K. Agbeti$^1$,\\ Arjan P. Palstra$^1$, Edwin Foekema$^1$, Leonel Aguilar$^2$, Ricardo da Silva Torres$^1$, Jascha Gr\"ubel$^{1, 2}$}
\IEEEauthorblockA{\textit{$^1$Wageningen University and Research}, Wageningen, The Netherlands \\
$^2$\textit{ETH Zurich}, Zurich, Switzerland}
}

% \author{
% \IEEEauthorblockN{Ali Youssef}
% \IEEEauthorblockA{\textit{Wageningen University}\\
% Wageningen, The Netherlands \\
% 0000-0002-9986-5324}
% \and
% \IEEEauthorblockN{Kristina Vodorezova}
% \IEEEauthorblockA{\textit{Wageningen University}\\
% Wageningen, The Netherlands \\
% 0000-0003-3518-2572}
% \and
% \IEEEauthorblockN{Yannick Aarts}
% \IEEEauthorblockA{\textit{Wageningen University}\\
% Wageningen, The Netherlands \\
% 0009-0000-6756-9136}
% \and
% \IEEEauthorblockN{Edwin Foekema}
% \IEEEauthorblockA{\textit{Wageningen University}\\
% Wageningen, The Netherlands \\
% 0000-0002-8746-5569}
% \and
% \IEEEauthorblockN{Wisdom E. K. Agbeti}
% \IEEEauthorblockA{\textit{Wageningen University}\\
% Wageningen, The Netherlands \\
% 0000-0002-8575-5351}
% \and
% \IEEEauthorblockN{Arjan P. Palstra}
% \IEEEauthorblockA{\textit{Wageningen University}\\
% Wageningen, The Netherlands \\
% 0000-0003-1470-6787}
% \and
% \IEEEauthorblockN{Leonel Aguilar}
% \IEEEauthorblockA{\textit{ETH Zurich}\\
% Zurich, Switzerland \\
% 0000-0001-6864-4492}
% \and
% \IEEEauthorblockN{Ricardo da Silva Torres}
% \IEEEauthorblockA{\textit{Wageningen University}\\
% Wageningen, The Netherlands \\
% 0000-0001-9772-263X}
% \and
% \IEEEauthorblockN{Jascha Gr\"ubel}
% \IEEEauthorblockA{\textit{Wageningen University \& ETH Zurich}\\
% Wageningen, The Netherlands \& Zurich, Switzerland \\
% 0000-0002-6428-4685}
% }

\maketitle

\begin{abstract}
IUMENTA (Latin for livestock) is an innovative software framework designed to construct and simulate digital twins of animals.
By leveraging the powerful capability of the Open Digital Twin Platform (ODTP) alongside advanced software sensors, IUMENTA offers researchers a user-friendly tool to seamlessly develop adaptive digital replicas of animal-based processes. 
This framework establishes a dynamic ecosystem that integrates insights from diverse experiments, consequently enhancing our understanding of animal behavioural and physiological responses. 
Through real-time tracking of an animal's energy balance.
IUMENTA provides valuable insights into metabolic rates, nutritional needs, emotional states, and overall well-being of animals. 
In this article, we explore the application of the IUMENTA framework in developing a digital twin focused on the animal’s energy balance. 
IUMENTA includes the EnergyTag system, a state-of-the-art wearable software sensor, which facilitates real-time monitoring of energy expenditure, allowing for continuous updates and personalisation of the energy balance digital twin.
\end{abstract}


\section{Introduction}

\begin{figure}[!t]
    \centering
    \includegraphics[width=\linewidth]{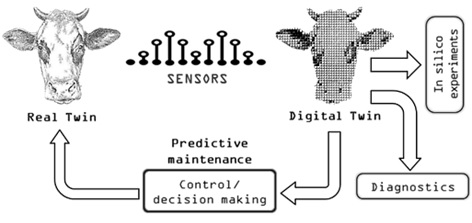}
    \caption{Illustration depicts the animal digital twins (ADT) concept. Sensors collect data from the real twin (animal-based system) to update the digital replica (twin) of the system. ADT can be utilized for predictive maintenance to support decision-making, as a diagnostic tool for early detection of problems, and in silico experiments and what-if simulations.}
    \label{fig:ADT_Scheme}
\end{figure}

Animal Digital Twins (ADTs) as a virtual representation of ``physical twins''~\cite{grubel2022hitchhiker}–--an individual animal, an underlying biological process, or a group of animals--- have the potential to rapidly evolve how animal research is conducted and evaluated (Fig.~\ref{fig:ADT_Scheme}).
This matters because on the one hand animal research has improved significantly to our understanding of various biological processes and thereby contributed scientific advancements and medical breakthroughs. 
It also has direct impact beyond human health and extends to agriculture and food security. 

On the other hand, animal research raises ethical concerns about the treatment and welfare of animals involved in experiments and scientific testing. 
This gives rise to the animal research ethical dilemma~\cite{Baumans2004}, which revolves around the tension between the potential benefits to humans and animals through scientific advancements and the moral responsibility to treat animals with respect and minimise their suffering.
In 2010, the European Parliament passed \emph{Directive 2010/63} on protecting animals used for scientific purposes, which explicitly requires scientists to consider the 3Rs (i.e., Replacement, Reduction, and Refinement) in all their work. 
The 3Rs framework~\cite{Tannenbaum2015} is a guiding principle in animal research ethics aimed at minimising the use of animals in scientific experiments and improving their welfare. 

By leveraging the rapid advancements in sensor and computational technologies, researchers are better equipped to adopt the 3Rs framework and address the ethical concerns tied to animal research. 
Numerous studies have demonstrated that high-fidelity computational models, or \textit{in silico} tools, such as E-Cell models~\cite{Tomita1999}, can accurately simulate a wide range of physiological processes. 
These advanced models often effectively can replace certain types of \textit{in vivo} testing, reducing the reliance on animal subjects and thus upholding the principles of Replacement~\cite{Tannenbaum2015}. 
The integration of these advanced technologies not only enhances the accuracy and efficiency of scientific research but also significantly contributes to the ethical goal of minimising animal use.

Constructing ADTs has three major benfits for ethical animal research and livestock management. First, ADTs provide a more accurate picture of the animals state than is typically maintained allowing for individual level research and management. Second, ADTs improve collective management and understanding by enabling more efficient resource allocation. Lastly, ADTs allow for long-term observation and understanding of the animals' genetics.

First, unlike conventional simulation models, ADT is continuously updated by incorporating data from sensors, providing a subject-specific and accurate depiction of the animal's individual physiological and behavioural responses with less invasive practices of acquiring the data.
This dynamic representation allows for personalised insights and interventions, ultimately leading to improved animal welfare and productivity.
The ADT can be utilised for a range of applications (Fig.~\ref{fig:ADT_Scheme}), including \textit{in silico} experiments (trials) to simulate the efficacy and safety of drug candidates or medical/veterinarian devices~\cite{Moingeon2023}. Additionally, digital twins significantly enhance diagnostics for early problem detection and enable predictive healthcare to foresee health and welfare issues.
Furthermore, digital twins enhance decision-making processes for farmers and farm managers. 
By integrating data from individual animals into a comprehensive farm management system, digital twins provide valuable insights into herd dynamics, environmental conditions, and overall farm performance~\cite{verdouw2021digital}. 
This enables more informed decisions, resulting in improved operational efficiency, productivity, and at the same time animal welfare.

Second, the implementation of digital twins also helps to optimise the allocation of resources~\cite{norton2019precision}. 
By accurately modelling and predicting the individual needs of each animal, in applications like precision livestock farming (PLF) ~\cite{Wathes2008,berckmans2017general, norton2019precision},  digital twins can ensure that resources, such as feed, water, and energy, are allocated with precision. 
This targeted approach not only reduces waste but also guarantees that each animal receives the appropriate nutrition and care tailored to its specific needs~\cite{verdouw2021digital}. 
For example, precision feeding systems can optimise nutrient supply based on an animal's growth rate, health status, and productivity goals, leading to better growth rates and improve overall health status~\cite{berckmans2017general}. 

Third, this holistic view enables more informed and strategic decisions, from breeding programs to herd health management and environmental sustainability practices. 
The predictive capabilities of digital twins also aid in planning and optimising breeding schedules, improving genetic selection, and ultimately enhancing the overall productivity and profitability of the livestock operation~\cite{verdouw2021digital}. 

However, how to implement ADTs faces several difficulties from generalisibility to maintenance.
Digital twins are often unique artefacts that cannot be reproduced or reused~\cite{grubel2022computer}. Especially in PLF, this can cause unnecessary redundancies as a unique digital twin is developed for each animal (type) by a diverse set of stakeholders.
Instead, we propose to produce reusable digital twins based on the Open Digital Twin Platform (ODTP)~\cite{grubel2023outlining}.
ODTP facilitates the combination of sensor networks, distributed cloud computing, loosely coupled micro-services, data semantics, containerisation, and automated visualisation to produce the infrastructure for ADTs~\cite{grubel2023ch}.
The digital revolution has highlighted the need for advanced sensor technologies capable of conducting complex computational inferences and estimating variables that cannot be directly measured with conventional hardware sensors~\cite{Youssef2023}. 
This need has led to the emergence of the software-sensing approach. 
Software sensors are fundamental to the Animal Digital Twin (ADT) concept, as they play a crucial role in inferring essential biological variables that are difficult to measure directly. 
By combining data from traditional hardware sensors with robust estimation models (algorithms), software sensors enable the ADT to provide a precise, real-time representation of animal's physiological and behavioural responses. 

\begin{figure*}[htbp]
\centerline{\includegraphics{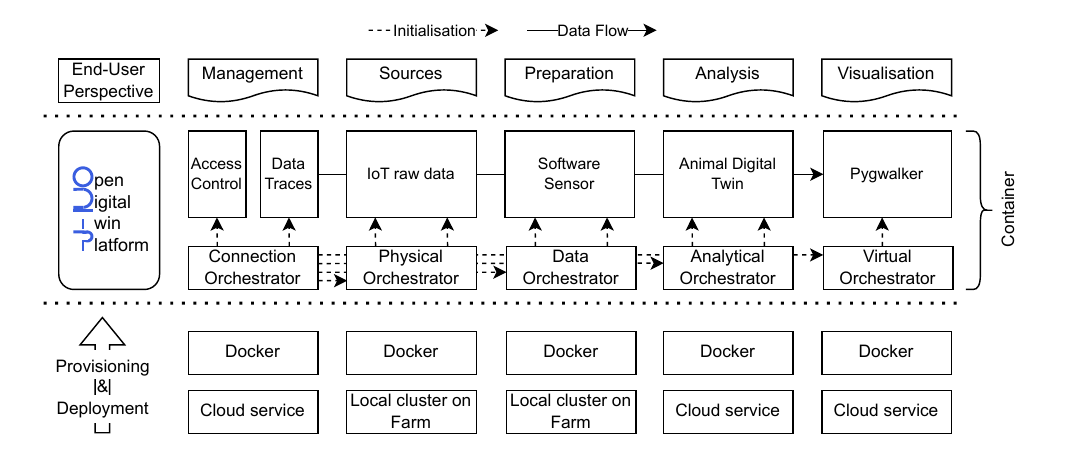}}
\caption{The IUMENTA framework embedded in the Open Digital Twin Platform (ODTP). ODTP runs on a server in the cloud managing the digital twin. Locally, at the farm, ODTP maintains two services to receive the raw data and instantiate the software sensors. The data is then transferred to the cloud and the animal digital twin computes the energy expenditure and all data is visualised in a service running the python visualisation library pygwalker.}
\label{fig:architecture}
\end{figure*}

We build a proof-of-concept prototype of a animal digital twin for energy expenditure based on ODTP and the EnergyTag software sensor, see Fig.~\ref{fig:architecture},  and demonstrate with three use cases that it can be validated for different animal species.

\section{Background}

\begin{figure}[!t]
    \centering
    \includegraphics[width=\linewidth]{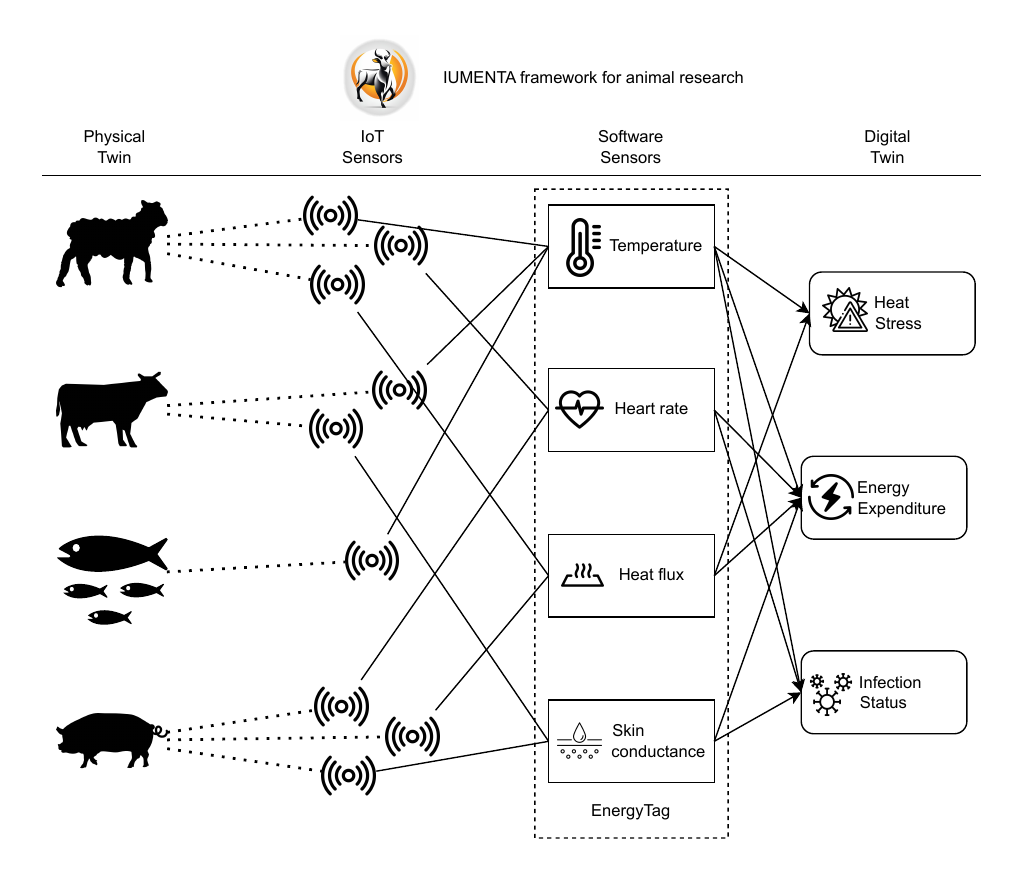}
    \caption{Software sensors facilitate the processing of complex animal data from raw sensors into semantically structured information. The software-defined sensors form the basis for the input in the energy expenditure calculation in the digital twin. Icons used with permission from flaticon authored by Freepik, kornkun, Octopocto, and Andrejs Kirma.}
    \label{fig:software:sensor}
\end{figure}

\subsection{Software sensors}

software sensor\cite{Youssef2023} combines data from hardware sensors with estimation algorithms or models to produce a measurement of a quantity that is otherwise not easily detectable, see Fig.~\ref{fig:software:sensor}.
For instance, a significant challenge in studying and monitoring living organisms is the continuous measurement of vital biological variables~\cite{Youssef2019}, such as animal metabolic rate and energy expenditure. 
These measurements are often hindered by a number of practical challenges fromphysical inaccessibility to invasiveness, the high cost, and unreliability of existing hardware sensors. 
To address these challenges, an auxiliary measurable variables can be introduced that indirectly estimates these otherwise immeasurable or inaccessible process variables. Instead of relying on direct physical measurements, the quantity needs to be computed in software..

The concept of software sensors~\cite{Youssef2023}, also referred to as inferential and virtual sensors~\cite{Tham1991}, is based on the idea of estimating vital biological variables—such as dynamic energy expenditure (DEE), heat stress, and infection state—that.
These variables are typically difficult to measure directly but can be reconstructed by integrating data obtained from one or more simple hardware sensors, like temperature and heart rate sensors, with robust estimation models. 
The combination of hardware and software enables the inference of complex biological information that would otherwise be very challenging, costly, and invasive to obtain.

\subsection{Digital twinning}

Digital twins represent a tangible object, process, or system by capturing key properties of a physical twin~\cite{grubel2022hitchhiker}.
These \emph{practical} digital twins may not attain similitude~\cite{glaessgen2012digital} but by focusing on specific aspects, such as the energy expenditure of animals, can sufficiently represent the physical twin for the task~\cite{grubel2022hitchhiker}.
Ultimately, digital twins only approach the real system of the physical twin, and the gaps to reality must be minded to ensure functionality~\cite{korenhof2021steering}.
Digital twins have been widely adopted across disciplines and a more critical approach may help to gain value from them~\cite{grubel2022hitchhiker,PURCELL2023101252}.
At the same time, a complex discourse is ongoing that questions what constitutes a digital twin and where to draw the line to simple systems such as digital models or digital shadows \cite{grubel2022hitchhiker}.
We consider practical digital twins as task-specific software systems that represent a physical twin sufficiently for a goal while leaving quantities that are not of interest unobserved.

Digital twins moved from industrial applications to agriculture in a bid to standardise farm procedures and increase utility~\cite{verdouw2021digital,Pylianidis2021COMPAG,Neethirajan2021Animals,FORE2024108676, doi:10.1080/27685241.2022.2150571,PURCELL2023100094}.
In particular, digital twins have been created for animals, crop fields, greenhouses, and farm buildings~\cite{verdouw2021digital,Pylianidis2021COMPAG,Neethirajan2021Animals,FORE2024108676, doi:10.1080/27685241.2022.2150571,PURCELL2023100094}.
In particular, ADTs have been used for monitoring the feeding behavior of dairy cows~\cite{ZHANG2023108181}; fish farming including metric estimation, and feeding, environmental, and health monitoring~\cite{UBINA2023100285}; and analyzing the energy demands in pig houses~\cite{10246265}.
Critically, like digital twins applications across the spectrum, these digital twins seem to be focused on a particular animal or location with visible efforts to generalise the software infrastructure.

The generation of digital twins is often still a piecemeal approach where components are stitched together manually to form an artifact \cite{grubel2022computer}.
In contrast, there is an increasing interest in creating digital twins in an automated fashion~\cite{grubel2023outlining, talasila2023dtaas}.
We focus on the Open Digital Twin Platform (ODTP) because of its generic architecture and the continuous support~\cite{grubel2023outlining}. 
ODTP focuses on modular components that are combined as a loose micro-service architecture.
In principle, ODTP supports distributed execution of digital twins allowing for computation to take place at an appropriate location at the edge (e.g. on a farm) or in the cloud~\cite{grubel2022hitchhiker}.
ODTP relies on the Open Digital Twin Workflow Standard (ODTWS) to provide interoperability to other digital twin platforms such as Digital Twins as a Service~\cite{talasila2023dtaas}.

\subsection{Experiments as Code}

ADTs can be used for monitoring, optimising, or gaining insights about animals, and this usually happens through the process of modelling and experimentation.
Often, multiple steps within an experiment are only informally encoded and hard to replicate~\cite{grubel2022computer}.
Digital twins for animal research also have the advantage that they allow us to formalise Experiments as Code~\cite{aguilar2024experiments} and thereby increase the reproducibility and reusability of animal research.

\begin{figure*}[htbp!]
    \centering
    \includegraphics[width=\linewidth]{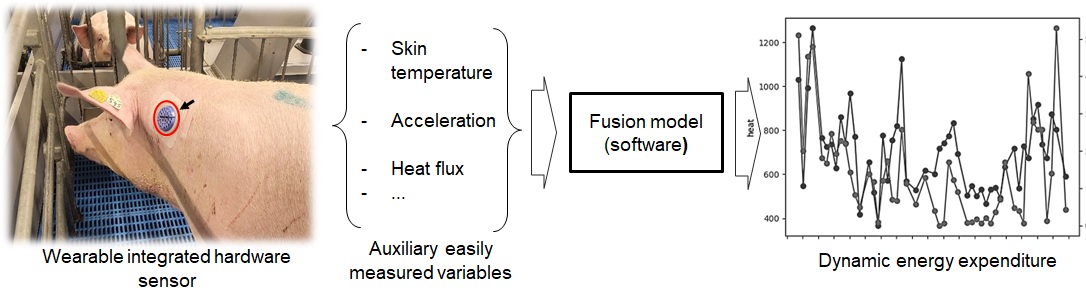 }
    \caption{The EnergyTag: A software Sensor for real-time monitoring of pig's dynamic energy expenditure. The EnergyTag comprises wearable integrated sensors (indicated by a red circle) designed to measure a number of auxiliary variables, including skin temperature, skin heat flux, ODBA (Overall Dynamic Body Acceleration), and heart rate. A fusion model (software) integrates sensor data to estimate the dynamic energy expenditure of animals.}
    \label{fig:EnergyTag}
\end{figure*}

Experiments as Code (ExaC)~\cite{aguilar2024experiments}is a paradigm that represents every component of an experiment—such as procedures, documentation, and infrastructure—in a structured, executable digital/code format. 
This concept allows experiments to be reproducible, auditable, reusable, debuggable, and scalable as all aspects of an experiment, from setup to execution and analysis, are documented and automated in code. 
This enables researchers to precisely replicate the experiment without ambiguities and reduces human errors in recreating or scaling it.
ExaC relies on automation to deploy experiments, and ADTs provide the ability to automate the representation and manipulation of animals in experiments, reducing the need for manual intervention. 

ADTs use real-time data from sensors, which can be processed and coded within the experiment framework, allowing for precise tracking of variables like metabolic rates or behaviour changes. 
Reproducibility, auditabibility, reusability, debuggability, and scalablility in the context of experimentation and experimental components are highly desirable if not essential for animal experiments. 
Regarding, Reproducibility and Scalability, ADTs ensure that an animal's dynamic and complex behaviors are captured digitally, potentially enabling experiments to be rerun or scaled without requiring physical animals for each test. 
Auditability enables third parties to ensure that the experiment complies with all the ethical and technical requirements, e.g., that the highest standards of animal care and wellbeing are maintained, or that the adequate model is applied for the data conversion of a specific animal or conditions. 
Reusability in Experiments as Code does not limit itself to reusing the animal samples/data but also reusing the components of the experiment itself, e.g., procedures, software, and hardware configuration.
Also, as with every digital representation, it becomes possible to debug an experiment, e.g., confirm if outliers are the result of a wrong experimental procedure, an atypical animal, or faulty sensors.

Formalising all steps within a digital twin also improves the accountability of research by making all steps from data acquisition to value extraction visible.
In the context of ADTs, implementing ExaC allows to follow the 3Rs framework more stringently while improving the accuracy of data and veracity of claims.

\section{Architecture of IUMENTA}

IUEMENTA focuses on data processing through software sensors and machine-learning models for animal digital twins, see Fig.~\ref{fig:architecture}. For generic digital twin functionality, the framework relies on ODTP. 
IUEMENTA takes a modular approach and created an ODTP component for each task that can be shared and reused. 
All code repositories and ODTP templates are available on our GitHub organisation, at \url{https://github.com/iumenta} (As of Oct. 2024).

\subsection{The EnergyTag software sensor}

Living organisms, including animals, are high-maintenance systems that require a constant supply of energy to sustain life. Consequently, they continuously expend energy, even when at rest, to support vital physiological processes, such as maintaining ion gradients, protein synthesis, and DNA repair. 
This energy expenditure (EE), or the rate at which an animal consumes energy, can fluctuate dynamically due to several internal and external factors. 
These factors include various physiological mechanisms (e.g., respiration and heart rates), behavioural activities (e.g., foraging), and psychological states (e.g., feelings of distress, including hunger) that collectively regulate and maintain the animal's energy balance. 
By monitoring the dynamic energy balance of animals, we can gain valuable insights into their metabolic rate, nutrient needs, affective states, and overall well-being.

Several methods have been established to measure EE, with indirect calorimetry in controlled environments considered the `gold standard'. 
This method assesses gas exchange between animals and their surroundings, where animals oxidise consumed nutrients by utilising oxygen ($VO_2$) and expelling carbon dioxide ($VCO_2$), thereby altering the gas composition within a respiration chamber. 
For unconstrained settings, the doubly labelled water method (DLW) estimates EE in free-ranging animals by administering isotopic water and measuring isotope excretion rates via urine, which allows for the calculation of $CO_2$ production and EE~\cite{Butler2004}. 
However, these standard methods are often unsuitable for real-time monitoring of free-ranging animals due to high costs, invasiveness, and impracticality for estimating short-term EE.

\begin{table*}[htbp!]
    \centering
    \caption{Overview of IUMENTA pipeline components}
    \begin{tabular}{lp{0.15\linewidth}p{0.15\linewidth}p{0.15\linewidth}p{0.2\linewidth}p{0.15\linewidth}}
         \toprule
         & \multicolumn{5}{c}{Components}\\
         \cmidrule(lr){2-6}
         &\textbf{Merge Sources }&\textbf{$ \rightarrow$ Quality Check }&\textbf{[$\rightarrow$ Split] }&\textbf{$\rightarrow$ Model (Training/Predict) }&\textbf{$ \rightarrow$ Report Generation}  \\
         \midrule
         Input & Multiple CSV files from sensors,
and a JSON file containing user-defined parameters. & The merged CSV file from the sensors and respira-
tion chamber. & The quality-checked CSV file and user-defined in-
structions in JSON format. & Training and test set CSV files, along with a JSON
file containing the model parameters.&Model parameters, predicted values, and instructions
from the JSON file.\\
         Output & A merged CSV file, combining the different raw
sensor data. & A quality-checked CSV file. & Separate training and test set CSV files.& A trained machine learning model with predicted
values and tuned parameters.& A PDF report containing the model’s performance
metrics and guidelines for future use.\\
         \bottomrule
    \end{tabular}
    \label{tab:iuemnta:components}
\end{table*}

In this context, a software sensing approach offers a promising solution by indirectly estimating EE through a range of easily measured variables. 
A notable advancement in this field is the development of a 
software sensor called the EnergyTag~\cite{aarts2024energytag} on top of wearable hardware sensors. 
The setup is designed for real-time monitoring of an animal’s dynamic energy expenditure (DEE). 
This innovative technology potentially overcomes the limitations of traditional methods by providing continuous, non-invasive, cost-effective monitoring in real-world settings.

The EnergyTag consists of wearable integrated hardware sensors (Fig.~\ref{fig:EnergyTag}) designed to measure an array of easily measured (auxiliary) variables, including skin temperature, skin heat flux, overall dynamic body acceleration (ODBA), and heart rate. 
These measurements are continuously collected at a frequency of one sample per second (1Hz) and are integrated using a fusion model (software) as illustrated in Fig.~\ref{fig:EnergyTag}. 
This real-time data integration enables precise estimation of dynamic energy expenditure (EE). 
For ease of use only wearables are considered that features onboard data storage capabilities with an option to broadcast the acquired data on demand to a client device, such as a smartphone, for further analysis and monitoring.

\subsection{Machine Learning Libraries}

To construct (indirect) measurements of complex biological variables, machine learning is applied to the raw data gathered from simple hardware sensors.
In IUMENTA, the data is automatically split into train and test data based on user specifications. 
Currently, IUMENTA supports a base version of prediction using linear regression and random forests. 
These modules are exchangeable and based on ODTP's component format \cite{grubel_2024_13843004}. 
The ODTP zoo repositories~\cite{grubel_2024_13843004} are used to share different models.
Depending on the research goals, different components may be re-combined from the zoo to quickly assemble pipelines.

%\begin{table}[htbp]
%\caption{Table Type Styles}
%\begin{center}
%\begin{tabular}{|c|c|c|c|}
%\hline
%\textbf{Table}&\multicolumn{3}{|c|}{\textbf{Table Column Head}} \\
%\cline{2-4} 
%\textbf{Head} & \textbf{\textit{Table column subhead}}& \textbf{\textit{Subhead}}& \textbf{\textit{Subhead}} \\
%\hline
%copy& More table copy$^{\mathrm{a}}$& &  \\
%\hline
%\multicolumn{4}{l}{$^{\mathrm{a}}$Sample of a Table footnote.}
%\end{tabular}
%\label{tab1}
%\end{center}
%\end{table}

\subsection{Pipelines}

The pipeline to develop digital twins for animal energy expenditure is designed to be modular. 
This means it consists of multiple components, some of which are mandatory, while others are optional, depending on the specific use case. 
This modular structure allows for flexibility and adaptability across different experimental setups and animal species.

A typical sequence of components in the pipeline is as follows depending on whether the model is being trained or used for prediction, see Tab.~\ref{tab:iuemnta:components} for component details.
Each step in the pipeline is interconnected, where the output of one component serves as part of the input for the next according to the ODTWS~\cite{grubel_2024_13843004}.
User instructions, typically provided in JSON format, are also integrated to guide the process at various stages. 
The modularity is further enhanced by the ability to run components interchangeably, allowing for the application of different machine-learning models. 
In our experiments with pigs and salmon, we employed both multivariate linear regression and random forest models.
This pipeline structure enables users to apply the trained models to real-time sensor data, potentially eliminating the need for controlled experimental conditions like those required in a respiration chamber.

In the following, we introduce the new ODTP components that form the IUEMENTA framework.
The input output mapping is shown in Table~\ref{tab:iuemnta:components}

\subsubsection{Merge Sources}

This component integrates data from multiple sources, such as sensors and respiration chambers, into a single CSV file. 
One of the main challenges addressed here is the synchronization of data with differing sampling rates. 
To tackle this, the pipeline either upsamples the less frequent time series or downsamples the more frequent ones. 
Both approaches have inherent drawbacks: downsampling may result in data loss while upsampling can introduce noise or distort the original data distribution. 

\subsubsection{Quality Check}

In this step, the data undergoes various quality control procedures, such as the identification of outliers and the handling of missing data. 
Users are provided with visual aids, including boxplots, line plots, and histograms, to assess the quality of each variable in the dataset.

\subsubsection{Split (optional)}

This step only applies when training the model. The processed dataset is divided into training and testing sets, which are used in the subsequent model training phase.

\begin{table*}
    \caption{Hardware wearable sensors used for the EnergyTag software sensor.}
    \centering
    \begin{tabular}{ccccc}
\toprule
        Device & Communication &Measurement & Accuracy & Sampling \\
\midrule
        \multirow{4}{*}{EnergyTag/CALERA\cite{aarts2024energytag}} & \multirow{4}{*}{Bluetooth\cite{etienne2023free}}& Heatflux & 0.5°C (1$\sigma$) \cite{verdel2021reliability}& 1Hz\\
        &&Skin temperature & ±0.05°C (typical) or ±0.13°C (max) from 20°C to 42°C & 1Hz\\
        && Core temperature & ± 0.28°C (1$\sigma$)\cite{zahner2023sensor} & 1Hz\\
        && Acceleration & not specified& 1Hz\\
        \midrule
        \multirow{1}{*}{Shimmer3 (Heart rate \cite{Lao2012})} & Bluetooth & PPG &  & 128 Hz\\
        \midrule
        \multirow{2}{*}{Thelma Biotel AT-LP7} & \multirow{2}{*}{Bluetooth}& Acceleration (ODBA) & range \(0 - 4.9035 \,{m/s}^2\); resolution  \( \pm 0.0136\, {m/s}^2\) & 25Hz  \\
        &&Oxygen consumption (MO\textsubscript{2}) & formula calculation \cite{biology13060393}& 1Hz\\
        
        \midrule
        \multirow{1}{*}{Aquadect Mosselmonitor} & Bluetooth& Bivalve movement \cite{kramer2001musselmonitor} & not specified & 0.5-1Hz \\
        \midrule
        \multirow{1}{*}{Oxygen Microsensor PM-PSt7 } & Wired & Respiration rate & $\pm$0.03\% O\textsubscript{2} saturation& 0.33Hz \\
        \midrule
        \multirow{1}{*}{ElectricBlue “Pulse V2”} & WiFi& Heart rate & Vishay CNY70 & 5-25Hz \\
\bottomrule
    \end{tabular}
    
    \label{tab:sensors}
\end{table*}

\subsubsection{Model Training/Prediction}

During this phase, a machine learning model is used to analyse the data. The Model is either being trained on the training dataset to predict energy expenditure on the test dataset for validation or uses the incoming data for continuous prediction on a previous user-provided training. 

\subsubsection{Report Generation}

The final step involves generating a report on the animal's energy expenditure and in case of traning it documents the model’s performance, including key metrics such as accuracy and precision. 
The report also provides details about the model’s parameters, making it easier to apply the trained model to new datasets in future applications.

\section{Use Cases}

IUMENTA has been developed as a generalisation for EnergyTag soft sensor to be applicable more easily to new animals and hardware sensors. 
To demonstrate the wide range of potential animal twins, we will now discuss the several cases were we are applying the EnergyTag and IUMENTA.
Currently, we cooperate with researchers to use our technology stack with pigs, salmon, and shellfish.
The described use cases focus on experiments with validation and are about training the ML model that can later be reused with new animals of the same kind.
For the different animals, different hardware sensors were used that get unified through the software sensor, see Table~\ref{tab:sensors}

\subsection{Pig Digital Twins}

\begin{figure}[!t]
    \centering
    \includegraphics[width=\linewidth,trim={13cm 5cm 13cm 10cm},clip]{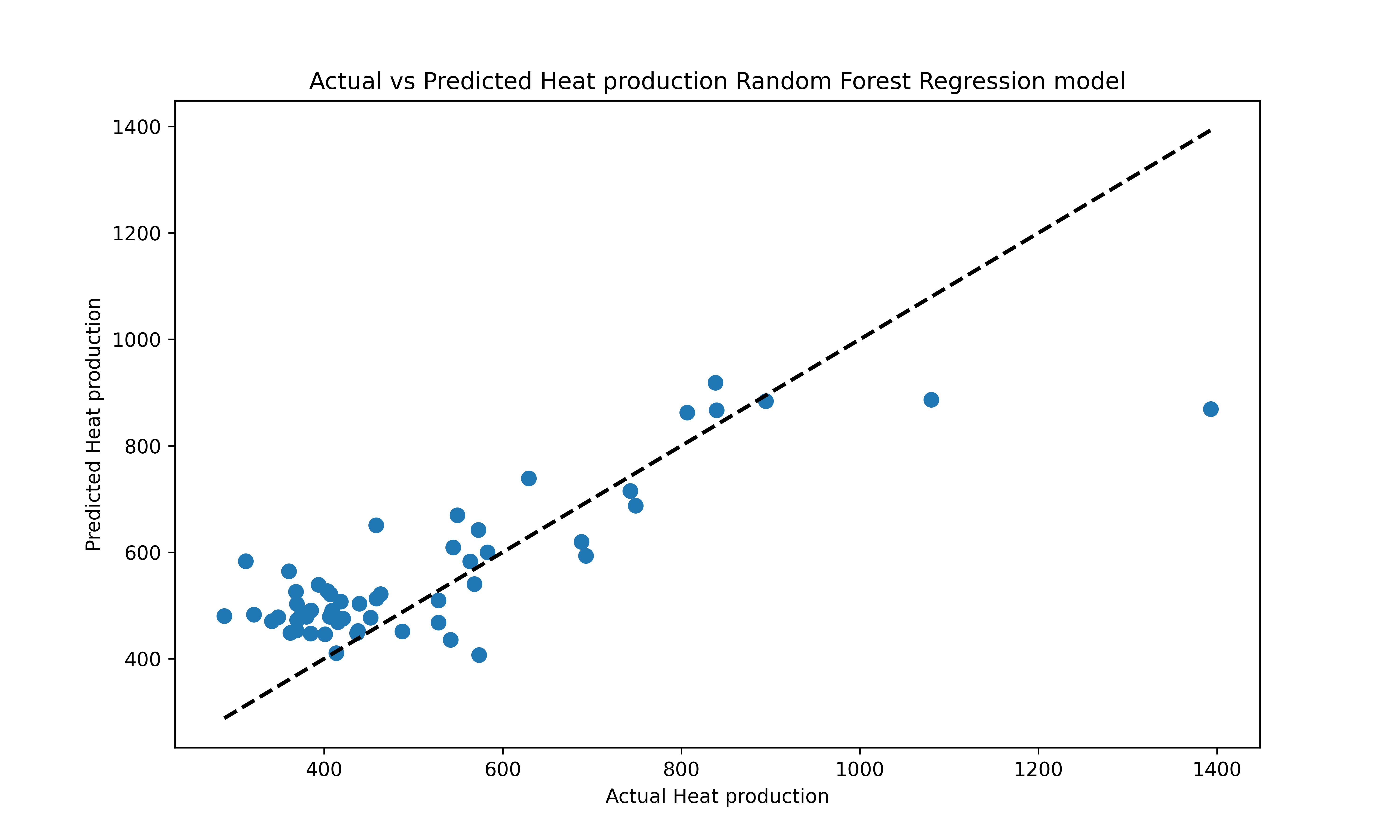}
    \caption{A random forest predicting the heat production of pigs based on the ``EnergyTag'' compared to a baseline~\cite{aarts2024energytag}. Image used with permission.}
    \label{fig:pig:prediction:pig}
\end{figure}

The first example ADT looks at heat production in pigs as a crucial measure to infer energy expenditure, see Fig.~\ref{fig:pig:prediction:pig}.

Previous work has demonstrated that heat flux can predict energy expenditure in humans\cite{lyden2014estimating, Youssef2019}.
Also, the acceleration of an animal was used to predict their energy expenditure~\cite{halsey2008acceleration}.
Therefore, we reuse the different sensors to improve our capacity to estimate energy expenditure.

The pipeline was applied to an experiment designed to predict the energy expenditure of pigs. 
In this experiment, the temperature inside a respiration chamber was manipulated to observe and record the pigs' heat expenditure under three distinct conditions: below thermoneutral (12 degrees Celsius), thermoneutral (22 degrees Celsius), and above thermoneutral (32 degrees Celsius). 
Heat was measured every three minutes, and data from sensors, external temperature, and respiration were recorded simultaneously. 
The sensors operated at varying sampling rates, which were reconciled during the data merging process. 
The wearable integrated sensor, which measures heat flux, skin temperature and acceleration samples at 1Hz and was used as the main predictor. 
For each 3 minute interval of heat production measurements, features where calculated for the continuously measuring sensors (Table~\ref{tab:sensors}). 

For this personalised model, the early measurements at a young age of a single pig are used to predict energy expenditure in the future, the last recorded measurements, corresponding to a roughly 5 to 1 train-test ratio.
Once a sufficiently accurate model is developed, it can be applied to sensor data alone, eliminating the need for a respiration chamber to estimate energy expenditure in future experiments.

\subsection{Salmon Digital Twins}

\begin{figure}[!t]
    \centering
    \includegraphics[width=\linewidth]{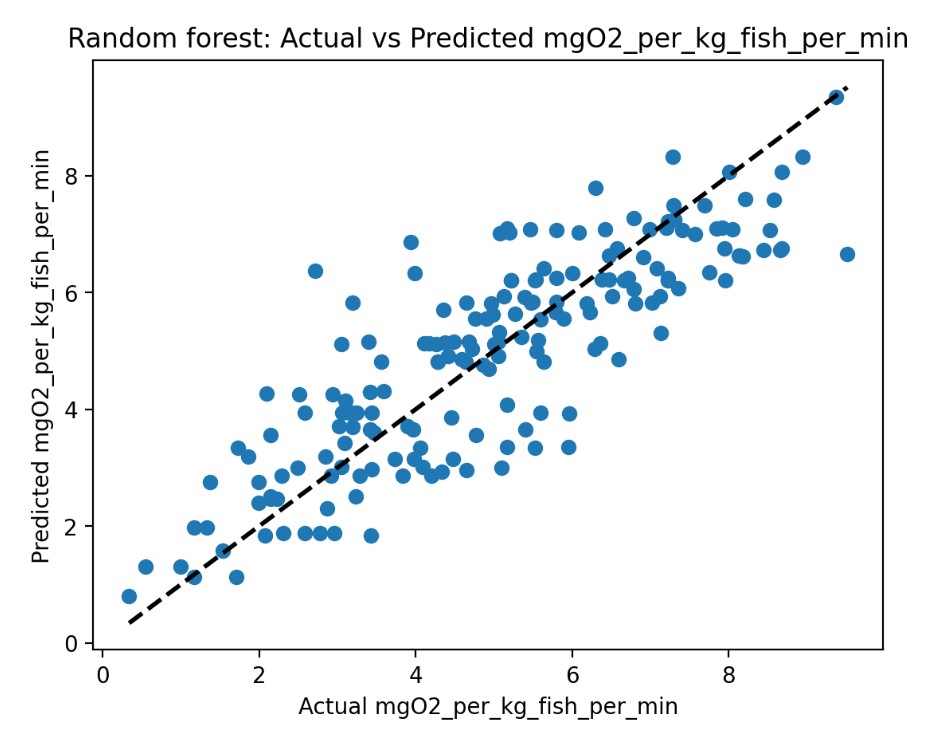}
    \caption{A random forest predicting the MO\textsubscript{2} of salmon based on the ODBA measurements compared to a real values of MO\textsubscript{2}}
    \label{fig:pig:prediction:salmon}
\end{figure}

The second example ADT looks at energy expenditure in salmon, see Fig.~\ref{fig:pig:prediction:salmon}.
In this case, energy expenditure of a salmon can be determined with the estimation of MO\textsubscript{2} from activity measured by an implantable sensor. Data was collected from swim tunnels with different unsteady flow speeds where the fish swam with the sensor measuring their 3-axial acceleration~\cite{biology13060393}. 
The acceleration data was recorded at a frequency of 25 Hz. 
The acceleration data was then transmitted to the receiver at an average interval of 40 s. 
This was the time interval used for each acceleration vs MO\textsubscript{2} calculation. 
Energy expenditure was measured in terms of oxygen consumption (mg O\textsubscript{2} per kilogram of fish weight), and acceleration was quantified using Overall Dynamic Body Acceleration (ODBA), a metric calculated using a proprietary formula. 
MO\textsubscript{2} was calculated from the oxygen concentration in the swim tunnel sampled every 1 second.

%Since the data had already been pre-processed—outliers had been removed, and the accelerometer data was filtered to eliminate saturation effects—the first two steps of the pipeline (Merge Sources and Quality Check) were omitted. 
%The remaining steps (Split, Model Training, and Report Generation) were used to complete the analysis.

\subsection{Shellfish Digital Twins}

\begin{figure}[t!]
    \centering
    \includegraphics[width=\linewidth,trim={0 0 0 1cm},clip]{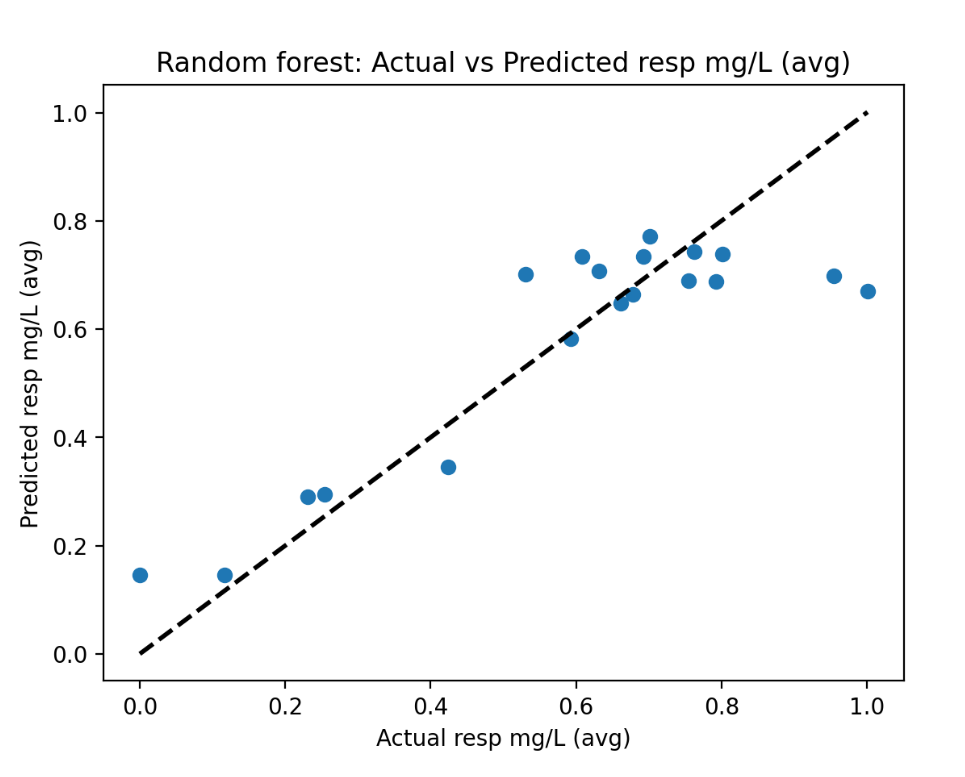}
    \caption{A random forest predicting the respiration rate of mussels based on the shell opening size and heart rate}
    \label{fig:pig:prediction:shellfish}
\end{figure}

The third example ADT looks at heat production in shellfish, see Fig.~\ref{fig:pig:prediction:shellfish}.
The modular design of the pipeline allows it to be adapted for different species and experiments, including the estimation of shellfish energy balance. 
Similar to the previous use cases, the pipeline components can be adjusted to match the specific requirements of the experiment, demonstrating the flexibility of this approach.
As the newest example, the available data is limited and we focus on the prediction of the respiration rate to provide a simple demonstration with mussels.

The data for the animal digital twin came from an experiment which investigates physiological responses in mussels exposed to increasing concentrations of SDS (sodium dodecyl sulfate). 
In this experiment, behavioral and physiological data were collected to assess how 2 individual mussels react to a chemical stressor. 
Specifically, shell opening, respiration, and heart rate were monitored over a 90-minute exposure period. 
Shell opening was measured in mm, recorded continuously to gauge the mussels' behavioral response. 
Respiration rate was determined through dissolved oxygen (DO) concentration in mg/L, representing metabolic activity, and heart rate was recorded in Beats Per Minute (BPM) to assess physiological stress levels. 
Data collection was time-stamped, with elapsed time recorded at 1-minute intervals.
This combined dataset provides insights into how increasing SDS concentrations influence both external activity and internal physiology in mussels. 
The ADT here helps to predict the metabolic activity of the mussels based on the size of the shell opening and the heart rate.

\section{Discussion}

Our example ADTs showcase the IUMENTA framework as a valid approach for automating animal research.
We adapted the general-purpose ODTP framework to create, manage, share and operate digital twins and have  successfully adapted it for the use in animal research. 
Whereas our proof-of-concept ADTs allow automated inference, we have not yet handled the integration of larger groups of animals.

Arbitrary deployment of IUMENTA is still under development. We also still need to work on the automation when other hardware sensors are used to ensure that IUMENTA can swiftly integrate new hardware sensors into the software sensor. At the same time, IUMENTA needs to handle distributed installation locally near the animal and remotely in the cloud better. 
We aim to integrate KAFKA-based data streams to improve the ability of software sensors to be formed \emph{ad hoc} and enable IUMENTA-based software sensors that are assembled on the fly.

The modular nature of this pipeline allows for adaptable and efficient data processing across a variety of animal species and experimental designs, while the implementation of Experiments as Code provides reproducibility and reusability. 
Its flexibility in model selection and the ability to skip or rearrange components make it a versatile tool in the development of digital twins for energy expenditure estimation. 
By enabling real-time analysis based on sensor data, this pipeline has the potential to transform energy expenditure modelling, reducing the need for traditional experimental setups like respiration chambers and improving the adoption of the 3Rs framework. 
Future research will continue to expand on these findings, applying the pipeline to a broader range of species and environmental conditions.

\section{Conclusion}
Digital twins in precision agriculture and biology in general and precision livestock and animal research in particular play a crucial role in implementing AI/ML systems.
We have demonstrated an end-to-end pipeline for animal research---IUMENTA. 
However, our platform has not only promise in animal research on the animal's energy expenditure but also to push the frontiers in real-time farming applications such as dispensing food.
By providing a ``laboratory in the wild''~\cite{grubel2022computer}, we can investigate animal behaviour on a previously unattainable level while also improving the animals' wellfare.

Digital twins are currently high on Gartner's hype curve and ultimately have to been seen through a more sombre lens to be applicable. 
However, by focusing digital twins on a particular task such as estimating energy expenditure, we demonstrate their usefulness in animal research.

\section*{Acknowledgment}

J.G. thanks Carlos Vivar Rios for the support with ODTP.
This project is part of the innovation programme ``Next Level Animal Science''(NLAS) of Wageningen University \& Research and also supported by \emph{Nederlandse
Organisatie voor Wetenschappelijk Onderzoek} through the grant ``Artificial Intelligence for Sustainable Food Systems'' 
(Nwa.1332.20.002).
All experimental protocols comply with the current laws of The Netherlands. 

For data pig data, the protocols were
approved by the Central Committee for Animal Experiments (CCD), project number
AVD1040020187184 and by the Animal Experiments Committee (DEC)
and Authority for Animal Welfare (IvD), experiment number 2022.W-0031.001 of Wageningen University.

For data salmon data, the protocols were
approved by the CCD, project number
AVD401002016652 and by the DEC
and IvD, experiment number 2016.D-0039.005 of Wageningen University. 
Salmon data was raised in the PhD project of W.E.K. Agbeti that was part of the research programme ``‘Whakapōhewa ki ahumoana - Reimagining Aquaculture'' (C11X1903) led by The New Zealand Institute for Plant and Food Research and funded by New Zealand’s Ministry of Business, Innovation and Employment Endeavour Fund.

No formal ethical conditions apply for work with shellfish.

\bibliographystyle{IEEEtran}
\bibliography{references,nordark,journal-articles}

\end{document}